\documentclass[1p,times,number]{elsarticle}
\usepackage[utf8]{inputenc}
\usepackage[T1]{fontenc}
\usepackage[greek,english]{babel}
\usepackage{alphabeta}
\usepackage{xspace}
\usepackage{amssymb}
\usepackage{amsthm}
\usepackage{amsmath}
\usepackage{fancyvrb}
\usepackage{verbatim}
\usepackage{xcolor}
\usepackage{soul}
\usepackage{url}
\usepackage{lineno}
\usepackage[
colorlinks,linkcolor={blue},citecolor={blue},urlcolor={red}%
]{hyperref}

\newcounter{bla}

\allowdisplaybreaks[4]

\swapnumbers

\theoremstyle{definition}

\DefineVerbatimEnvironment{sverb}{Verbatim}{%
framesep=2mm,xleftmargin=\parindent}
\sethlcolor{white}
\newcommand{\cprime}{\/{\mathsurround=0pt$'$}}

\newcommand*{\pd}[2]{\mathchoice{\frac{\partial#1}{\partial#2}}
  {\partial#1/\partial#2}{\partial#1/\partial#2}
  {\partial#1/\partial#2}}

\newcommand*{\fd}[2]{\mathchoice{\frac{\delta#1}{\delta#2}}
  {\delta #1/\delta#2}{\delta#1/\delta#2}{\delta#1/\delta#2}}

\newcommand{\ddx}[1]{\partial_x^{#1}}

\newcommand{\eval}[2][\right]{\relax
  \ifx#1\right\relax \left.\fi#2#1\rvert}

\newcommand{\envert}[2][\right]{\relax
  \ifx#1\right\relax \left\lvert\else#1\lvert\fi#2#1\rvert}

\newcommand{\enVert}[2][\right]{\relax
  \ifx#1\right\relax \left\lVert\else#1\lVert\fi#2#1\rVert}

\let\kappa\varkappa
\let\phi\varphi

\newcommand{\cde}{CDE\xspace}

\newcommand{\reduce}{Reduce\xspace}

\providecommand{\href}[2]{#2}

\newcommand{\mc}[1]{\mathcal{#1}}
\newcommand{\mb}[1]{\mathbb{#1}}
\DeclareMathOperator{\Mat}{Mat}

\begin{document}

\begin{frontmatter}
\title{Weakly nonlocal Poisson brackets:
  \\
tools, examples, computations}

\author{M. Casati}
\address{School of Mathematics and Statistics
  \\
  Ningbo University, Ningbo, China}
\ead{matteo@nbu.edu.cn}

\author{P. Lorenzoni}
\address{Dipartimento di Matematica e Applicazioni,
  \\
  Universit\`a di Milano-Bicocca, Milano, Italy
  \\
  and Sezione INFN, Milano-Bicocca}
\ead{paolo.lorenzoni@unimib.it}

\author{D. Valeri}
\address{Dept. of Mathematics ``G.~Castelnuovo''
  \\
  Sapienza Universit\`a di Roma, Rome, Italy}
\ead{daniele.valeri@uniroma1.it}

\author{R. Vitolo}
\address{Dept.\ of Mathematics and Physics ``E. De Giorgi''
  \\
  Universit\`a del Salento, Lecce, Italy
  \\
  and Sezione INFN, Lecce
}
\ead{raffaele.vitolo@unisalento.it}
\ead[url]{http://poincare.unisalento.it/vitolo}

\begin{abstract}
  We implement an algorithm for the computation of Schouten bracket of weakly
  nonlocal Hamiltonian operators in three different computer algebra systems:
  Maple, Reduce and Mathematica. This class of Hamiltonian operators encompass
  almost all the examples coming from the theory of (1+1)-integrable
  evolutionary PDEs.

\end{abstract}

\end{frontmatter}

\section{Introduction}

Poisson brackets are ubiquitous in Mathematical and Theoretical Physics.
Besides their traditional role in Mechanics, they play an important role in the
study of integrable evolutionary systems of partial differential equations
(PDEs) in two independent variables $t$, $x$ and $n$ dependent variables
${\bf u}=(u^1,...,u^n)$
\begin{equation}
  \label{eq:12}
  u^i_t = f^i({\bf u},{\bf u}_x,{\bf u}_{xx},....),\qquad i=1,...,n,
\end{equation}
where $f^i$ are differential polynomials.  We refer to the books
\cite{ablowitz91:_solit,DubrovinKricheverNovikov:InSI,FT,NMPZ,Zakharov:WIsIn}
for an account of this theory. In this context a \emph{local} Poisson bracket
of two local functionals $F=\int f({\bf u},{\bf u}_x,{\bf u}_{xx},....)dx$ and
$G=\int g({\bf u},{\bf u}_x,{\bf u}_{xx},....)\, dx$ is a local functional of
the form
 \begin{equation}
  \label{eq:22}
  \{F,G\}_P = \int\fd{F}{u^i}P^{ij}\fd{G}{u^j}\,dx.
\end{equation}
where
\[P^{ij}=\sum_{\sigma}P^{ij}_{\sigma}({\bf u},{\bf u}_x,{\bf
    u}_{xx},....)\partial_x^{\sigma}
\]
is a suitable (matrix) differential operator and $\fd{F}{u^i}$ and
$\fd{G}{u^i}$ are the variational derivatives of the functionals $F$ and $G$
respectively. The differential polynomials $f$ and $g$ are called the
\emph{densities} (of the functionals $F$ and $G$).

The operator $P$ cannot be arbitrarily chosen. The fulfillment of skew-symmetry
and the Jacobi identity for the Poisson bracket imposes strict constraints on
$P$: it must be formally skew-adjoint and the Schouten bracket $[P,P]$ must be
identically zero (see \emph{e.g.}  \cite{Dorfman:DSInNEvEq}).\footnote{Explicit
  formulas for Schouten bracket will be given later.}

A system of evolutionary PDEs of the form \eqref{eq:12} is said to be
Hamiltonian w.r.t. the Hamiltonian operator $P$ if there exists a local
functional $H=\int h\,dx$, called the \emph{Hamiltonian functional} such that
\begin{equation}
  \label{eq:23}
  f^i({\bf u},{\bf u}_x,{\bf u}_{xx},....)= P^{ij}\fd{H}{u^j}
\end{equation}

The integrability of the system \eqref{eq:23} is revealed in the existence of
infinitely many local functionals (including $H$) in involution with respect to
the Poisson bracket defined by $P$.  A universal procedure to construct a
sequence $H_0=H,H_1,H_2,...$ of Hamiltonian functionals in involution is
Magri's bi-Hamiltonian recursion \cite{Magri:SMInHEq}:
\begin{equation}
  \label{eq:46}
   P^{ij}_1\fd{H_{k+1}}{u^j} =  P^{ij}_2\fd{H_{k}}{u^j}.
\end{equation}
It relies on the existence of a pair of Hamiltonian operators $(P_1,P_2)$ satisfying the compatibility condition
\[[P_1,P_2]=0,\]
where the square bracket  denotes  (as above) the Schouten bracket.

In general computation of the Schouten bracket is a highly nontrivial
computational task and most of the examples can be handled only with the help
of computer algebra systems. Recently in \cite{casati16:_master_walg,Vit19} two
computer algebra packages for calculations of Schouten bracket for \emph{local}
differential operators have been introduced and described. The packages have
been developed following two different mathematical approaches to differential
operators: the algebraic approach through Poisson Vertex Algebras
\cite{casati16:_master_walg} and the supermanifold approach \cite{Vit19}. A
fairly complete list of packages for local operators can be found in
\cite{Vit19}.

Since the beginning of research on integrable systems, it was clear that
Poisson brackets defined by local differential operators did not cover all
interesting examples. Nonlocal differential operators
(i.e. integro-differential operators) are much more common: they can be found
for the KdV equation, the modified KdV equation, the AKNS equation, the
Nonlinear Schr\"odinger equation, etc. (see \cite{Wan02}).

A systematic research on nonlocal operators was started by Ferapontov and
Mokhov \cite{FM90}, then widely generalized by Ferapontov in subsequent works
\cite{FerAMS}. A further generalization was proposed by Maltsev and Novikov
that introduced the general class of weakly nonlocal operators \cite{MN01}:
\begin{equation}
\label{eq:10}
P^{ij} =\sum_{\sigma}P^{ij}_{\sigma}({\bf u},{\bf u}_x,{\bf u}_{xx},....)\partial_x^{\sigma}+
\sum_{\alpha,\beta=1}^Ne^{\alpha\beta}w^i_{\alpha}({\bf u},{\bf u}_x,{\bf u}_{xx},....)\partial_x^{-1} w^j_{\beta}({\bf u},{\bf u}_x,{\bf u}_{xx},....),
\end{equation}
where the coefficients $P^{ij}_{\sigma}$, $w^i_\alpha$, $w^j_\beta$ are
differential polynomials, $e^{\alpha\beta}$ is a symmetric constant matrix and
the operator $\partial_x^{-1}$ is defined to be
\begin{equation}
\label{eq:11}
\partial_x^{-1} = \frac{1}{2}\int_{-\infty}^x\,dy -
\frac{1}{2}\int^{+\infty}_x\,dy.
\end{equation}
 All the above mentioned examples of  nonlocal
operators have this form, thus certifying the importance and generality of that class.

The computation of Schouten bracket for nonlocal operators is a more difficult
task with respect to the local case. Indeed, until recently it was not even
clear how to reach a divergence-free canonical form of the Schouten bracket in
the presence of nonlocal terms. Such canonical (or normal) form is crucial in
order to get a set of necessary and sufficient conditions for the vanishing of
the Schouten bracket.  It is worth to emphasize that the only publicly
available computer algebra package for nonlocal operators is the Maple package
\texttt{JET} \cite{meshkov02:_tools_pdes}. Its strongest limit is that it is
not always able to simplify the nonlocal terms because it lacks for an
algorithm to that purpose.

Recently, a canonical form of the Schouten bracket for weakly nonlocal
operators has been achieved in three different formalisms (distributions,
differential operators and Poisson Vertex Algebras) leading to three different
(but equivalent) algorithms for the computation of the Schouten bracket
\cite{CLV19} (see \cite{LV:SuperPoisson} for the supermanifold approach).

 The aim of this paper is to implement the algorithms of \cite{CLV19} in three
 different computer algebra systems, respectively, Maple, Reduce and
 Mathematica.  Let us describe more in detail such implementations.
\begin{itemize}
\item \emph{Maple implementation.} The package for Maple (called
  \texttt{Jacobi}) is the straightforward implementation of the algorithm
  introduced in \cite{CLV19} for the computation of Schouten bracket using the
  language of distributions.  The package provides the syntax and the command
  for computing the formula for the Schouten bracket given in \cite{DZ} and
  reducing it to the normal form described in \cite{L,CLV19}. We provide
    both a classical and a parallel implementation of the algorithm. We used
    Maple's \texttt{Grid} package and we split the computation in few parallel
    independent branches when possible. We found that strategy to be
    optimal as there is an overhead related with the initialization of the
    independent processes. While parallel computing has no effect on relatively
    simple computations (or makes things even worse), it reduces the timing to
    half when the calculation is more complex.
\item \emph{\reduce implementation.} This implementation is within the software
  package \cde\footnote{This is an acronym standing for Calculus on
    Differential Equations}, an official \reduce package \cite{reduce} devoted
  to calculations in the geometry of differential equations and integrable
  systems \cite{KVV17}.  Since version 2.0 (2015), \cde can quite easily
  compute the conditions for a nonlocal operator to be Hamiltonian for a given
  partial differential equation (in the sense of mapping conserved quantities
  into symmetries), see \cite{KVV17} for examples. It can also compute the
  Jacobi property for \emph{local} operators \cite{Vit19}. The newly added
  library \texttt{cde\_weaklynl.red} provides the syntax and the command for
  computing Schouten brackets of weakly nonlocal operators. Here, the command
  uses the \cde data structure for differential operators and \cde treatment of
  nonlocal variables in order to introduce the data and carry out the
  algorithm.

  At the moment, \texttt{Reduce} can be installed by means of a binary
    snapshot of the official distribution. Such snapshots can be found from the
    official website. The latest snapshot at the time of writing can be found
    at the following link:
  \url{https://sourceforge.net/projects/reduce-algebra/files/snapshot_2021-07-16/}
  and it contains all examples that are discussed in our paper. The main
    reason for using \texttt{Reduce} were the fact that it is free software, so
    that we have been able to learn how to code in an efficient way, and the
    fact that it has an extremely fast computer algebra engine, which is
    competitive with commercial software. The real drawbacks of \texttt{Reduce}
    at the moment are that it has no satisfactory Graphical User Interface and
    it has no parallel computing capabilities. However, a web player is
    currently being implemented at the official website.
\item \emph{Mathematica implementation.}  This implementation extends the
  package \texttt{MasterPVA.m}\footnote{The acronym is motivated from the fact
    that the package computes the so-called Master Formula (see
    \cite{BarakatSoleKac:PVAlTHEq}) in PVAs} \cite{casati16:_master_walg},
  devoted to the computation of $\lambda$-brackets in local PVAs, to the case
  of weakly non-local PVAs.  The new added package
  \texttt{nlPVA.wl} (developed for Mathematica 11) provides the
  syntax and the command for computing the three terms appearing in the Jacobi
  identity \eqref{eq:jacobi} for weakly non-local $\lambda$-brackets.
  The computation is done using the Master Formula for PVAs as in
  \cite{CLV19}, for the local part of the $\lambda$-bracket, and the algorithm
  in \cite{casati16:_master_walg} for the non-local part. In particular, the
  package can be used to compute \eqref{eq:jacobi} for the local case as
  well. We provide both a classical and a parallel implementation of the
    algorithm; the latter is available in the package \texttt{nlPVA-par.wl}
\end{itemize}
The three implementations are different in that the mathematical formalism
  on which each of them is based is different \emph{and} the computer algebra
  system on which each of them is based is different. Despite such
  differences, the canonical divergence-free form of the results can be
  translated from one formalism to another, see \cite{CLV19}. However, the
  three implementations follow different paths in order to achieve
  corresponding results. That is due to the different data structure that they
  have to handle (distributions, differential operators, Poisson Vertex
  Algebras).

The software packages and examples are released under the terms of the FreeBSD
license. They can be downloaded at the webpage of this paper in the Geometry of
Differential EQuations website, the URL is in \cite{gdeq}.

\paragraph{Structure of the paper}
For each implementation we will shortly recall its mathematical procedure.
Then, we will discuss the details of the corresponding software package. For
each software package we will present the same three examples of (systems of)
partial differential equations that can be written in Hamiltonian form through
a weakly nonlocal operator, and compute the Schouten bracket of any such
operator.

The examples are: the \emph{modified KdV} equation, the \emph{Heisenberg
  magnet} system, for which we will also compute the compatibility of its
weakly nonlocal operator with a second local Hamiltonian operator, and one new
example of weakly nonlocal operator for a \emph{WDVV equation} in the form of a
quasilinear first-order system of PDEs. Also the examples can be
downloaded from the web page of this paper at \cite{gdeq}.

We observe that examples with non-zero output are not available. Indeed,
  there are no known examples of an evolutionary PDE in two independent
  variables which have two Hamiltonian operators that are not compatible. For
  some PDEs that depends on the fact that there exists no non-trivial
  three-vectors, provided that their linearization fulfills a certain property
  (see \cite{KKV04}). It is also easy to check that simple changes in input
  (\emph{e.g.} in a sign or in a coefficient) yield non-zero results, showing
  the `rigidity' of the results.

\section{Maple implementation}
\label{sec:maple-implementation}

In what follows, let us denote the independent variable by $x$, the dependent
variables by $(u^i)$, $1\leq i\leq n$, and derivatives by
$u^i_\sigma=\pd{^\sigma u^i}{x^\sigma}$; note that if $\sigma=0$ then
$u^i_\sigma = u^i$.  The \emph{total derivative} $\partial_x$ is
defined as
\begin{equation}\label{eq:8}
  \partial_x = \pd{}{x} + u^i_{\sigma+1}\pd{}{u^i_\sigma},
\end{equation}
(the summation convention holds as usual).

The Schouten bracket of weakly nonlocal Poisson bivectors is implemented in the
Maple package \texttt{Jacobi} (filename \texttt{jacobi.mpl}) using the language
of distributions (see, e.g. \cite{CLV19,DZ}). Let us briefly describe the
mathematical algorithm. We will need three different labels for the independent
variable: $x$, $y$, $z$. The $\sigma$-th power of $\partial_x$ is denoted by
$\partial_x^\sigma$, and analogously for other variables.

A weakly nonlocal Poisson bivector has the form
\[
  P^{ij}_{x,y}=
  \sum_{k\ge 0}B^{ij}_k(u^h,u^h_\sigma)\delta^{(k)}(x-y)+ c^{\alpha\beta} w^i_{\alpha}(u^k,u^k_\sigma)
  \nu(x-y)w^j_{\beta}(u^k,u^k_\sigma)
\]
where $\nu(x-y)=\frac{1}{2}\text{sgn}(x-y)$. In the above formula it is understood that
\[P^{ij}_{x,y}=-P^{ji}_{y,x}\]
This implies $c^{\alpha\beta}=c^{\beta\alpha}$.

Following  \cite{DZ} the Schouten bracket of  two weakly nonlocal Poisson bivectors $P^{ij}_{x,y}$ and
\[ Q^{ij}_{x,y}=
  \sum_{k\ge 0}C^{ij}_k(u^h,u^h_\sigma)\delta^{(k)}(x-y)+ d^{\alpha\beta} z^i_{\alpha}(u^k,u^k_\sigma)
  \nu(x-y)z^j_{\beta}(u^k,u^k_\sigma)\]
can be defined as
\begin{equation}\label{eq:20}
  \begin{split}
[P,Q]^{ijk}_{x,y,z}=&\frac{\partial P^{ij}_{x,y}}{\partial u^{l}_\sigma(x)}
\partial_x^\sigma Q^{lk}_{x,z}+ \frac{\partial P^{ij}_{x,y}}{\partial u^{l}_\sigma(y)}
\partial_y^\sigma Q^{lk}_{y,z}+\frac{\partial P^{ki}_{z,x}}{\partial u^{l}_\sigma(z)} \partial_z^\sigma
Q^{lj}_{z,y}+\\
&\frac{\partial P^{ki}_{z,x}}{\partial u^{l}_\sigma(x)}
\partial_x^\sigma Q^{lj}_{x,y}
+\frac{\partial P^{jk}_{y,z}}{\partial u^{l}_\sigma(y)}
\partial_y^\sigma Q^{li}_{y,x}
+ \frac{\partial P^{jk}_{y,z}}{\partial u^{l}_\sigma(z)}
\partial_z^\sigma Q^{li}_{z,x}+\\
&\frac{\partial Q^{ij}_{x,y}}{\partial u^{l}_\sigma(x)}
\partial_x^\sigma P^{lk}_{x,z}
+ \frac{\partial Q^{ij}_{x,y}}{\partial u^{l}_\sigma(y)}
\partial_y^\sigma P^{lk}_{y,z}
+\frac{\partial Q^{ki}_{z,x}}{\partial u^{l}_\sigma(z)} \partial_z^\sigma
P^{lj}_{z,y}+\\
&\frac{\partial Q^{ki}_{z,x}}{\partial u^{l}_\sigma(x)}
\partial_x^\sigma P^{lj}_{x,y}
+\frac{\partial Q^{jk}_{y,z}}{\partial u^{l}_\sigma(y)}
\partial_y^\sigma P^{li}_{y,x}
+ \frac{\partial Q^{jk}_{y,z}}{\partial u^{l}_\sigma(z)}
\partial_z^\sigma P^{li}_{z,x}
\end{split}
\end{equation}
The vanishing of the Schouten bracket $[P,Q]^{ijk}_{x,y,z}$
 means that for any choice of the test functions
$p_i(x),q_j(y),r_k(z)$ the triple integral
\begin{equation}
  \iiint  [P,Q]^{ijk}_{x,y,z} p_i(x)q_j(y)r_k(z)\,dxdydz
\end{equation}
should vanish. Using distributional identities it is possible to reduce the Schouten bracket
 to a normal form \cite{CLV19,L}:
\begin{enumerate}
\item Using the identities
  \begin{equation}
\begin{split}
\nu(z-y)\delta(z-x)&=\nu(x-y)\delta(x-z)\\
\nu(y-x)\delta(y-z)&=\nu(z-x)\delta(z-y)\\
\nu(x-z)\delta(x-y)&=\nu(y-z)\delta(y-x)
\end{split}\label{eq:15}
\end{equation}
and their differential consequences we can eliminate all terms containing
 $\nu(z-y)\delta^{(n)}(z-x)$, $\nu(y-x)\delta^{(n)}(y-z)$, $\nu(x-z)\delta^{(n)}(x-y)$ producing nonlocal terms containing
 $\nu(x-y)\delta^{(n)}(x-z)$, $\nu(z-x)\delta^{(n)}(z-y)$, $\nu(y-z)\delta^{(n)}(y-x)$ and additional local terms.
\item  Using the identities
  \begin{equation}
\begin{split}
f(z)\delta^{(n)}(x-z)&=\sum_{k=0}^n\binom{n}{k}f^{(n-k)}(x)\delta^{(n-k)}(x-z),\\
f(y)\delta^{(n)}(z-y)&=\sum_{k=0}^n\binom{n}{k}f^{(n-k)}(z)\delta^{(n-k)}(z-y), \\
f(x)\delta^{(n)}(y-x)&=\sum_{k=0}^n\binom{n}{k}f^{(n-k)}(y)\delta^{(n-k)}(y-x),
\end{split}	 \label{eq:16}
\end{equation}
we can eliminate the dependence on $z$  in the coefficients of the terms containing $\nu(x-y)\delta^{(n)}(x-z)$, the dependence on $y$  in the coefficients of the terms containing $\nu(z-x)\delta^{(n)}(z-y)$ and the dependence on $x$  in the coefficients of the terms containing $\nu(y-z)\delta^{(n)}(y-x)$.
 After the first two steps the Schouten bracket has the form
\begin{multline}
  c_1(x,y,z)\nu(x-y)\nu(x-z)+c_2(x,y,z)\nu(y-z)\nu(y-x)
  \\
  +c_3(x,y,z)\nu(z-x)\nu(z-y)+ \sum_{n\ge 0}a_n(x,y)\nu(x-y)\delta^{(n)}(x-z)
  \\
  +\sum_{n\ge 0}b_n(x,z)\nu(z-x)\delta^{(n)}(z-y)+
  \sum_{n\ge 0}c_n(y,z)\nu(y-z)\delta^{(n)}(y-x)+\dots
\end{multline}
where $\dots$ are local terms.
\item The remaining local part can be reduced to the form
  \begin{equation}
    \sum_{m,n}f_{mn}(x)\delta^{(m)}(x-y)\delta^{(n)}(x-z)\label{eq:14}
  \end{equation}
using the identities (and their differential consequences)
  \begin{equation}
    \delta(z-x)\delta(z-y)=\delta(y-x)\delta(y-z)=\delta(x-y)\delta(x-z)
    \label{eq:19}
\end{equation}
and the identities \eqref{eq:16}.
\end{enumerate}
As proved in \cite{CLV19} no further simplifications are possible and the
vanishing of the Schouten bracket turns out to be equivalent to the vanishing
of each coefficient in the reduced form.

We provided two implementations of the above algorithm: a classical,
  non-parallel implementation in the library file \texttt{jacobi.mpl} and
  an implementation that uses Maple parallel programming routines in the
  library file \texttt{jacobi\_p.mpl}. The difference in the execution
  times of the examples is discussed below, as well as some details of the
  implementation of the parallel code. The most important thing to be
  remembered when using the parallel code is that it \emph{requires} $4$ cores
  to run correctly. On a modern computer that is not a problem; however, errors
  could arise when trying to run the parallel code on an older version of
  Maple/older computer, or, for example, when running Maple on a virtual
  machine.

\subsection{The mKdV equation}
\label{sec:mkdv-equation-maple}

We consider the mKdV equation $u_t = u^3u_x + u_{xxx}$ and its nonlocal
Hamiltonian operator
\begin{equation}
  \label{eq:7}
  P = \partial_x^3 + \frac{2}{3}(u^2 \partial_x + uu_x - u_x \partial_x^{-1}u_x)
\end{equation}
(see, \emph{e.g.}, \cite{Wan02}). The Hamiltonian property of $P$ is proved by
checking that the Schouten bracket vanishes, $[P,P]=0$.

First of all, we set up the number of components
\begin{verbatim}
N:=1:
\end{verbatim}
and the maximal order of derivative coordinate $(u_x,u_{xx},\ldots)$ appearing in the coefficients of the operator $P$
\begin{verbatim}
M1:=1:
\end{verbatim}
and the maximal order of derivative of the $\delta$ function appearing in the operator $P$
\begin{verbatim}
D1:=3:
\end{verbatim}
Since the two arguments of the Schouten bracket coincide we define
\begin{verbatim}
M2:=M1:D2:=D1:
\end{verbatim}
The field variables $u^i_\sigma$ are introduced as \texttt{u[i,x,\sigma]} (it
is not possible to change the symbols \texttt{u} and \texttt{x}).  We introduce
the nonlocal part $w^i_j u^j_x \partial_x^{-1}w^k_h u^h_x$ of the operator,
$w^i_j$ is \texttt{W[i,j]} in the program:
\begin{verbatim}
W:=Matrix(N,N):
W[1,1]:=1:
\end{verbatim}
Finally, we input the differential operator as a $3$-dimensional array. The
last index must be $0$ in input; it will be increased throughout the
calculation.
\begin{verbatim}
P := Array(1..N,1..N,0..M2);
P[1, 1, 0] := delta[x - y, 3] + 2/3*u[1, x, 0]^2*delta[x - y, 1]
 + 2/3*u[1, x, 0]*u[1, x, 1]*delta[x - y, 0]
 - 2/3*W[1, 1]*u[1, x, 1]*delta[x - y, -1]
  *subs(x = y, W[1, 1])*u[1, y, 1];
\end{verbatim}
Again, the notation \texttt{delta} and the use of the second independent
variable \texttt{y} are not modifiable by the user.  \hl{Internally, the maximal
order of derivative coordinate is set to be equal to}
\begin{verbatim}
M:=M1+D1+M2+D2;
\end{verbatim}
The main procedure is
loaded by
\begin{verbatim}
read(`jacobi.mpl`);
\end{verbatim}
where the file \texttt{jacobi.mpl} shall be in the current Maple
directory. \hl{Alternatively, an implementation that makes some use of the
  parallel programming capabilities of Maple can be used by means of}
\begin{verbatim}
read(`jacobi_p.mpl`);
\end{verbatim}
The result of the calculation is a $3$-vector, and it shall be
defined beforehand:
\begin{verbatim}
T := Array(1..N,1..N,1..N):
\end{verbatim}
We can now call the main procedure; the output will be stored in $T$, or
$T=[P,P]$:
\begin{verbatim}
Schouten_bracket(P,P,T,N,M1,D1,M2,D2);
\end{verbatim}
The program starts computing, and prints the stages of the procedure as
follows. The first message
\begin{verbatim}
"Step 0: calculating Schouten bracket"
\end{verbatim}
is printed when the program is computing formula \eqref{eq:20}.  The iterated
total derivatives of the components of $P$ are calculated by induction up to
the order $M-1$, then the results are multiplied by the coefficients of $P$ and
summed up and internally stored in a $3$-vector \texttt{T0}. \hl{In the
  parallel version, since the calculation has to be repeated for $P$ and $Q$
  (if they are different), two parallel processes are run through the
  \texttt{Grid} library.} Then
\begin{verbatim}
"Step 1 of the algorithm"
\end{verbatim}
is showed when the identities \eqref{eq:15} are used in order to reduce the
nonlocal part. A short notation for the product of \texttt{delta} and a step
function is used:
\begin{verbatim}
delta[z-y,-1,z-x,0]:=delta[x-y,-1]*delta[x-z,0]:
\end{verbatim}
Here the result is stored in the $3$-vector \texttt{T1}. The
reduction is completed by means of the identities \eqref{eq:16} when the
message
\begin{verbatim}
"Step 2 of the algorithm"
\end{verbatim}
is issued, defining a new $3$-vector \texttt{T2} and completing the reduction
of the nonlocal part to the canonical form. \hl{Since the reduction process can
  be split into several independent calculations, Step 2 is also split into 4
  parallel calculations in} \texttt{jacobi\_p.mpl}.

The reduction is completed by means of the identities \eqref{eq:19} that act on
the local part only, after the message
\begin{verbatim}
"Step 3 of the algorithm"
\end{verbatim}
The final result is stored in the $3$-vector \texttt{T3} that is returned to
the command line and shown; $0$ in this case. The user can show all components
of the result one by one, instead of getting them altogether, by the command
\begin{verbatim}
for i to N do
  for j to N do
    for k to N do
      print(SB[i, j, k] = T[i, j, k]);
    end do;
  end do;
end do;
\end{verbatim}
\hl{The execution time is 0.09s for} \texttt{jacobi} \hl{and 0.31s for the
  parallel version} \texttt{jacobi\_p}, \hl{showing that the parallel
  programming model that we adopted has an overhead that makes its use in
  simple computations not so effective. The calculations for all Maple examples
  have been performed on a laptop with an Intel i7-8565U processor.}

\subsection{The Heisenberg magnet equation}
\label{sec:heis-magn-equat-maple}

The following operator is an example of nonlocal Hamiltonian
operator of a class introduced by Ferapontov (see \cite{FerAMS} and references
therein):
\begin{equation}
  \label{eq:101}
  P = f^2
  \begin{pmatrix}
    1 & 0\\ 0 & 1
  \end{pmatrix}\partial_x
  + f
  \begin{pmatrix}
    u^1 u^1_x + u^2 u^2_x & u^1 u^2_x - u^2 u^1_x
    \\
    u^2 u^1_x - u^1 u^2_x & u^1 u^1_x + u^2 u^2_x
  \end{pmatrix}
  +
  \begin{pmatrix}
    u^1_x \partial_x^{-1} u^1_x & u^1_x \partial_x^{-1} u^2_x
    \\
    u^2_x \partial_x^{-1} u^1_x & u^2_x \partial_x^{-1} u^2_x
  \end{pmatrix}
\end{equation}
where $f=(1/2)((u^1)^2 + (u^2)^2 + 1)$. The operator $P$ is a Hamiltonian
operator for the Heisenberg magnet equation \cite{FerAMS}:
\begin{align}
  \label{eq:13}
  &u^1_t = u^2_{xx} + \frac{1}{f}\big(u^2(u^1_x)^2 - 2u^1u^1_xu^2_x
    - u^2(u^2_x)^2\big),
  \\
  &u^2_t = - u^1_{xx} - \frac{1}{f}\big(u^1(u^2_x)^2 - 2u^2u^1_xu^2_x
    - u^1(u^1_x)^2\big).
\end{align}
The Schouten bracket $[P,P]$ is computed in Maple as follows. We introduce the
number of components and a maximal order of \hl{$x$-derivatives of the dependent variables $u^i$  and of the delta function}
\begin{verbatim}
N:=2;
M1:=1;
D1:=1,
M2:=M1;
D2:=D1;
\end{verbatim}
and we define the operator:
\begin{verbatim}
f := u[1,x,0]^2/2 + u[2,x,0]^2/2 + 1/2;
P := Array(1..N,1..N,0..M2);
P[1,1,0] := f^2*delta[x-y,1] + f*(u[1,x,0]*u[1,x,1]
 + u[2,x,0]*u[2,x,1])*delta[x-y,0] + u[1,x,1]*delta[x-y,-1]*u[1,y,1];
P[2,2,0] := f^2*delta[x-y,1] + f*(u[1,x,0]*u[1,x,1]
 + u[2,x,0]*u[2,x,1])*delta[x-y,0] + u[2,x,1]*delta[x-y,-1]*u[2,y,1];
P[1,2,0] := f*(u[1,x,0]*u[2,x,1]-u[2,x,0]*u[1,x,1])*delta[x-y,0]
 + u[1,x,1]*delta[x-y,-1]*u[2,y,1];
P[2,1,0] := f*(u[2,x,0]*u[1,x,1]-u[1,x,0]*u[2,x,1])*delta[x-y,0]
 + u[2,x,1]*delta[x-y,-1]*u[1,y,1];
\end{verbatim}
after loading the program, the bracket is computed as in the previous example:
\begin{verbatim}
read(`jacobi.mpl`);
T := Array(1..N,1..N,1..N);
Schouten_bracket(P,P,T,N,M1,D1,M2,D2);
\end{verbatim}
The components of the output $3$-vector can be shown one by one:
\begin{verbatim}
for i to N do
  for j to N do
    for k to N do
      print(SB[i,j,k] = T[i,j,k]);
    end do;
  end do;
end do;
SB[1, 1, 1] = 0
SB[1, 1, 2] = 0
SB[1, 2, 1] = 0
SB[1, 2, 2] = 0
SB[2, 1, 1] = 0
SB[2, 1, 2] = 0
SB[2, 2, 1] = 0
SB[2, 2, 2] = 0
\end{verbatim}
The Heisenberg magnet equation is indeed bi-Hamiltonian. The second Hamiltonian
operator can be written in the above coordinates as the \hl{ultra-local}
operator
\begin{equation}
  \label{eq:21}
  Q=f^2
  \begin{pmatrix}
    0 & -1 \\ 1 & 0
  \end{pmatrix}\delta(x-y).
\end{equation}
\hl{In this case $M_1=D_1=M_2=D_2=0$.}
The Hamiltonian property is easily checked. First of all we load the operator:
\begin{verbatim}
Q := Array(1..N,1..N,0..M):
Q[1,1,0] := 0;
Q[2,2,0] := 0;
Q[1,2,0] := -f^2*delta[x-y,0];
Q[2,1,0] := f^2*delta[x-y,0];
\end{verbatim}
Then we compute the Schouten bracket:
\begin{verbatim}
S := Array(1..N,1..N,1..N);
Schouten_bracket(Q,Q,S,N,M1.D1,M2,D2);
\end{verbatim}
which yields zero. Finally, we calculate the compatibility condition between
the two Poisson brackets:
\begin{verbatim}
U := Array(1..N,1..N,1..N);
Schouten_bracket(P,Q,U,N,M1,D1,M2,D2),
\end{verbatim}
\hl{where $M_1=1,D_1=1,M_2=0,D_2=0$}.
The compatibility, in this case, has a nice geometric interpretation: the
Hamiltonian operator $Q$ turns out to be a Killing--Poisson tensor with respect
to the metric $(g_{ij})$ that forms the leading coefficient of the Hamiltonian
operator $P$ (see \cite{mokhov98:_sympl_poiss}).

\hl{Here, the execution time for the Schouten bracket $[P,P]$ is 0.48s for}
\texttt{jacobi} \hl{and 0.54s for the parallel version}
\texttt{jacobi\_p}. \hl{This is the most complicated between the
brackets that we calculate for this example, although still not
challenging. For a significant difference between the non-parallel and the
parallel version, see the next example.}

\subsection{The equations of associativity}
\label{sec:equat-assoc-maple}

In \cite{Dub2dQFT} the geometric theory of the equations of associativity, or
Witten--Dijkgraaf--Verlinde--Verlinde equation, is developed in detail. One of
the cases of equations of associativity can be rewritten as a quasilinear
system of PDEs of the form $u^i_t=(V^i(\mathbf{u}))_x$ (Example 3 in
\cite{FPV16}), with
\begin{equation}
  \label{eq:111}
  u^1_t = (u^2+u^3)_x,\quad
  u^2_t = \left(\frac{u^2(u^2+u^3) - 1}{u^1}\right)_x,\quad
  u^3_t = u^1_x.
\end{equation}
It can be proved that the above system admits the following operator of
Ferapontov type \cite{VV21}:
\begin{equation}
  \label{eq:37}
  P = g^{ij}\ddx{} + \Gamma^{ij}_{k}u^k_x
  +\alpha_1 \pd{V^i}{u^q}u^q_x\ddx{-1}\pd{V^j}{u^p}u^p_x
  +\gamma_1 u^i_x\ddx{-1}u^j_x,
\end{equation}
where the metric in \emph{upper} indices is
\begin{equation}\footnotesize
  \label{eq:38}
  g^{ij}=\begin{pmatrix}u_{1}^{2}
         +u_{2}^{2}
         +2  u_{2}  u_{3}
         +u_{3}^{2}
         +1&
         \frac{u_{1}^{2}  u_{2}
               +u_{2}^{3}
               +2  u_{2}^{2}  u_{3}
               +u_{2}  u_{3}^{2}
               -u_{2}
               -u_{3}}{
               u_{1}}&u_{1}
         (u_{2}
           +2  u_{3}
         )
         \cr
         \frac{u_{1}^{2}  u_{2}
               +u_{2}^{3}
               +2  u_{2}^{2}  u_{3}
               +u_{2}  u_{3}^{2}
               -u_{2}
               -u_{3}}{
               u_{1}}&
         \frac{u_{1}^{2}  u_{2}^{2}
               +4  u_{1}^{2}
               +u_{2}^{4}
               +2  u_{2}^{3}  u_{3}
               +u_{2}^{2}  u_{3}^{2}
               -2  u_{2}^{2}
               -2  u_{2}  u_{3}
               +1}{
               u_{1}^{2}}&u_{2}^{2}
         +2  u_{2}  u_{3}
         -3\cr
         u_{1}
         (u_{2}
           +2  u_{3}
         )
         &u_{2}^{2}
         +2  u_{2}  u_{3}
         -3&u_{1}^{2}
         +u_{3}^{2}
         +4
       \end{pmatrix}
\end{equation}
the Christoffel symbols are
\begin{equation}
       \label{eq:82}
       \Gamma^{ij}_1=\begin{pmatrix}u_{1}&u_{2}&u_{3}\cr
         \frac{-
               (u_{2}^{2}
                 +u_{2}  u_{3}
                 -1
               )

               (u_{2}
                 +u_{3}
               )
               }{
               u_{1}^{2}}&
         \frac{-
               (u_{2}^{2}
                 +u_{2}  u_{3}
                 -1
               )
               ^{2}}{
               u_{1}^{3}}&
         \frac{-
               (u_{2}^{2}
                 +u_{2}  u_{3}
                 -1
               )
               }{
               u_{1}}\cr
         u_{2}
         +u_{3}&
         \frac{u_{2}^{2}
               +u_{2}  u_{3}
               -1}{
               u_{1}}&u_{1}
        \end{pmatrix}
\end{equation}
\begin{equation}
       \label{eq:83}
       \Gamma^{ij}_2 =\begin{pmatrix}u_{2}
         +u_{3}&
         \frac{u_{2}^{2}
               +u_{2}  u_{3}
               -1}{
               u_{1}}&u_{1}\cr
         \frac{(2  u_{2}
                 +u_{3}
               )
               (u_{2}
                 +u_{3}
               )
               +u_{1}^{2}}{
               u_{1}}&
         \frac{2  u_{2}^{3}
               +3  u_{2}^{2}  u_{3}
               -u_{3}
               +u_{1}^{2}  u_{2}
               +
               (u_{3}^{2}
                 -2
               )
                 u_{2}}{
               u_{1}^{2}}&2
         (u_{2}
           +u_{3}
         )
         \cr
         0&0&0
         \end{pmatrix}
\end{equation}
\begin{equation}
         \label{eq:84}
         \Gamma^{ij}_3=\begin{pmatrix}u_{2}
         +u_{3}&
         \frac{u_{2}^{2}
               +u_{2}  u_{3}
               -1}{
               u_{1}}&u_{1}\cr
         \frac{(u_{2}
                 +u_{3}
               )
                 u_{2}}{
               u_{1}}&
         \frac{(u_{2}^{2}
                 +u_{2}  u_{3}
                 -1
               )
                 u_{2}}{
               u_{1}^{2}}&u_{2}\cr
         u_{1}&u_{2}&u_{3}
         \end{pmatrix}
\end{equation}
and the value of the constants is $\alpha_1=-1$, $\gamma_1=-1$.

In Maple, we need to introduce the coefficients as three arrays
\begin{verbatim}
g := Matrix(N,N);
g[1,1] := u[1,x,0]^2 + (u[2,x,0] + u[3,x,0])^2 + 1;
\end{verbatim}
and so on, and
\begin{verbatim}
Gamma := Array(1..N,1..N,1..N);
Gamma[1,1,1] := u[1,x,0];
\end{verbatim}
and so on, and finally
\begin{verbatim}
W := Matrix(N,N);
W[1,1] := 0;
W[1,2] := 1;
W[1,3] := 1;
W[2,1] := -(u[2, x, 0]*(u[2, x, 0] + u[3, x, 0]) - 1)/u[1, x, 0]^2;
W[2,2] := (2*u[2, x, 0] + u[3, x, 0])/u[1, x, 0];
W[2,3] := u[2, x, 0]/u[1, x, 0];
W[3,1] := 1;
W[3,2] := 0;
W[3,3] := 0;
\end{verbatim}
\hl{where \texttt{W[i,j]} is equal to $\pd{V^i}{u^j}$.  Indeed, in this example
  the nonlocal part is made by two matrices, the other being the
  identity.} Then, we define the operator as
\begin{verbatim}
P := Array(1..N,1..N,0..M);
for i to N do
  for j to N do
    P[i,j,0] := g[i,j]*delta[x - y,1]
      + add(Gamma[i,j,k]*u[k,x,1],k = 1..N)*delta[x - y,0]
      - add(add(W[i,s]*u[s,x,1]*delta[x - y,-1]
       *subs(x = y,W[j,t])*u[t,y,1],s = 1..N),t = 1..N)
      - u[i,x,1]*delta[x - y,-1]*u[j,y,1];
  end do;
end do;
\end{verbatim}
The Schouten bracket is computed as usual and the result is $0$.

\hl{Here, the execution time for the Schouten bracket $[P,P]$ is about 18s for}
\texttt{jacobi} \hl{and about 9s for the parallel version} \texttt{jacobi\_p},
\hl{showing that the parallel calculations are giving an effective reduction
  with respect to the non-parallel implementation.}

\section{\reduce implementation}
\label{sec:impl-cdiff-cde}

Now, we describe the implementation of the Schouten bracket for nonlocal
differential operators in \reduce, as a part of the package \cde.  The package
\cde has been developed by one of us (RV) for calculations in the geometric
theory of PDEs and integrability \cite{KVV17}.

Consider two weakly nonlocal \textbf{skew-adjoint} operators $P$, $Q$:
\begin{align}\label{eq:18}
  & P(\psi)^i = P^{ij}\psi_j =
  B^{ij\sigma}\partial_\sigma\psi_j + c^{\alpha\beta}
  w^i_\alpha\partial_x^{-1}(w^j_\beta\psi_j),
  \\ \label{eq:29}
  & Q(\psi)^i = Q^{ij}\psi_j =
  C^{ij\sigma}\partial_\sigma\psi_j + d^{\alpha\beta}
  z^i_\alpha\partial_x^{-1}(z^j_\beta\psi_j),
\end{align}
For the nonlocal summands, the skew-adjointness is fulfilled if
$c^{\alpha\beta}$ and $d^{\alpha\beta}$ are symmetric matrices.

Let us introduce the \emph{nonlocal variables}
\begin{equation}
  \label{eq:2}
  \tilde{\psi}_\alpha^a = \partial_x^{-1}(w^i_\alpha\psi_i^a),\qquad
  \tilde{\chi}_\beta^a = \partial_x^{-1}(z^i_\beta\psi_i^a),
\end{equation}
where $a=1$, $2$, $3$ and $\alpha$ and $\beta$ run in the same finite set of
indices (see \cite{Many99,KVV17} for a geometric
theory of nonlocal variables).

We have the formula:
\begin{multline}
  \label{eq:17}
  \ell_{P,\psi}(\varphi)^i = \pd{B^{ij\sigma}}{u^k_\tau}\partial_\sigma\psi^1_j
  \partial_\tau \varphi^k +
  c^{\alpha\beta} \pd{w^{i}_\alpha}{u^k_\tau}\partial_\tau \varphi^k
  \partial_x^{-1}(w^j_\beta\psi_j)
  \\
  + c^{\alpha\beta} w^{i}_\alpha\partial_x^{-1}
  \left(
    \pd{w^j_\beta}{u^k_\tau}\partial_\tau \varphi^k\psi_j
  \right),
\end{multline}
where we used the fact that $\partial_x^{-1}$ commutes with
linearization. Then, we have the following expression for the Schouten bracket
\cite{Dorfman:DSInNEvEq}:
\begin{equation}
  \label{eq:283}
  \begin{split}
  [P,Q](\psi^1,\psi^2,\psi^3) = &[\ell_{P,\psi^1}(Q(\psi^2))(\psi^3)
                                  + \text{cyclic}(\psi^1,\psi^2,\psi^3)
  \\
  & +\ell_{Q,\psi^1}(P(\psi^2))(\psi^3)
  + \text{cyclic}(\psi^1,\psi^2,\psi^3)]
  \end{split}
\end{equation}
where square brackets stand for horizontal cohomology \cite{Many99}, i.e. the
expression should be considered up to total $x$-derivatives.

A single summand of the above formula has the expression
\begin{equation}
  \label{eq:3}
  \begin{split}
  \ell_{P,\psi^1}(Q(\psi^2))(\psi^3) = &
  \pd{B^{ij\sigma}}{u^k_\tau}\partial_\sigma\psi^1_j
  \partial_\tau\Big(C^{kp\sigma}\partial_\sigma\psi_p^2
  + d^{\alpha\beta} z^k_\alpha\tilde{\chi}^2_\beta\Big)\psi^3_i
  \\
   & + c^{\alpha\beta} \pd{w^{i}_\alpha}{u^k_\tau}\partial_\tau
   \Big(C^{kp\sigma}\partial_\sigma\psi_p^2
  + d^{\gamma\delta} z^k_\gamma\tilde{\chi}^2_\delta\Big)
  \tilde{\psi}^1_\beta\psi^3_i
  \\
   &   - c^{\alpha\beta}\tilde{\psi}^3_\alpha\Big(\pd{w^j_\beta}{u^k_\tau}
      \partial_\tau (
    C^{kp\sigma}\partial_\sigma\psi^2_p
    + d^{\gamma\delta} z^k_\gamma \tilde{\chi}^2_\delta)\psi_j^1\Big)
    \end{split}
\end{equation}
Here, the last summand has been obtained by integration by parts:
\begin{multline}
    \label{eq:1}
    c^{\alpha\beta} w^{i}_\alpha\partial_x^{-1}
    \left(\pd{w^j_\beta}{u^k_\tau}\partial_\tau (
    C^{kp\sigma}\partial_\sigma\psi^2_p
    + d^{\gamma\delta} z^k_\gamma \tilde{\chi}^2_\delta)^k\psi_j^1
    \right)\psi^3_i =
  \\
  - c^{\alpha\beta}\tilde{\psi}^3_\alpha\Big(\pd{w^j_\beta}{u^k_\tau}\partial_\tau (
    C^{kp\sigma}\partial_\sigma\psi^b_p
    + d^{\gamma\delta} z^k_\gamma \tilde{\chi}^2_\delta)\psi_j^1\Big) +
    \partial_x(T),
  \end{multline}
where the term $\partial_x(T)$ is ignored in the bracket formula.

In \reduce, the above formulae~\eqref{eq:283} and \eqref{eq:1} and the
subsequent algorithm is implemented through \cde in the library file
\texttt{cde\_weaklynl.red}. We illustrate the implementation and the usage by
means of the following program files, which provide the same computations as in
the previous Section. The examples accompany the paper as \reduce
program files. They can be run by the command
\begin{verbatim}
in "filename.red";
\end{verbatim}
issued as the first command in a \reduce window.

\subsection{The mKdV equation}
\label{sec:mkdv-equation}

The calculation is carried out with the operator $P$ in \eqref{eq:7}.
First of all, we load the package \cde:
\begin{sverb}
  load_package cde;
\end{sverb}
then we define the list of variables:
\begin{sverb}
  indep_var:={x};
  dep_var_equ:={u};
  loc_arg:={psi};
  total_order:=10;
\end{sverb}
Here \texttt{indep\_var} is the name of the independent variable,
\texttt{dep\_var\_equ} is the list of dependent variables that are involved in
the equation and \texttt{loc\_arg} is a list of symbols that will be used for
the local arguments of the three-vector (one symbol for each dependent
variable in the equation). Finally, \texttt{total\_order} is the maximal order
of derivative that can appear in all expressions that will be input in and/or
calculated by the program. \hl{The order can be computed more sharply with the
formula that has been given in the Maple package: \texttt{M:=M1+D1+M2+D2}, but
in any case CDE will issue an error and quit if the order turns out to be too
low.}

We load the operator $P$ as follows. First of all, we define the local part of
$P$ using the \cde syntax \cite{KVV17}:
\begin{sverb}
  mk_cdiffop(ham_l,1,{1},1);
  for all psi let ham_l(1,1,psi)=
    td(psi,x,3) + (2/3)*(u**2*td(psi,x) + u*u_x*psi);
\end{sverb}
Then we enter the nonlocal part by the commands
\begin{sverb}
  mk_wnlop(c,w,1);
  c(1,1):= - 2/3;
  w(1,1):=u_x;
\end{sverb}
The command \texttt{mk\_wnlop} defines the coefficient matrix of the nonlocal
summand $c^{\alpha\beta}$ as a \reduce operator \texttt{c}, where the value at
the indices \texttt{alpha}, \texttt{beta} is \texttt{c(alpha,beta)}.  It also
defines the vectors $w^i_\alpha$ (for every index $\alpha$) as the reduce
operator \texttt{w}, where the value at the indices \texttt{i}, \texttt{alpha}
is \texttt{w(i,alpha)}. The third entry of the command \texttt{mk\_wnlop} is
the number of summands $w^i_\alpha$ (for every index $i$), or the range of the
index $\alpha$.

The first weakly nonlocal operator is arranged in a list as follows:
\begin{verbatim}
ham1:={ham_l,c,w};
\end{verbatim}
Since we are computing $[P,P]$, the second operator is equal to the first one:
\begin{verbatim}
ham2:=ham1;
\end{verbatim}

The \reduce implementation makes use of nonlocal variables in order to
calculate the Schouten bracket. For this reason, we need to form a list of all
distinct non-local variables. There is one nonlocal variable for any value
of $\alpha$; in the mKdV case, the list has the format
\begin{verbatim}
nloc_var:={{tpsi,w,1}};
\end{verbatim}
The jet space (\emph{i.e.} all variables and their $x$-derivatives) is prepared
by the command
\begin{verbatim}
dep_var_tot:=cde_weaklynl(indep_var,dep_var_equ,loc_arg,nloc_var,
  total_order);
\end{verbatim}
creates the jet space with:
\begin{enumerate}
\item derivative symbols like \texttt{u\_x}, \texttt{u\_2x}, \dots;
\item three times the symbols in the list \texttt{loc\_arg},  by appending
  \texttt{1, 2, 3} to the names of the arguments. In our case: \texttt{psi1,
    psi2, psi3} and their derivatives \texttt{psi1\_x, psi2\_x, \dots}
\item three times the nonlocal variables in \texttt{nloc\_arg}, i.e.
  \texttt{tpsi\_11,tpsi\_12,tpsi\_13}, which correspond to the nonlocal
  variables $\tilde{\psi}^1_1$, $\tilde{\psi}^2_1$, $\tilde{\psi}^3_1$.
  \begin{sverb}
    tpsi1_x = w(1,1)*psi1;
    tpsi2_x = w(1,1)*psi2;
    tpsi3_x = w(1,1)*psi3;
  \end{sverb}
\end{enumerate}
The above jet space is defined using well-tested \cde routines, after an
initial consistency check on the input data (for example, the number of
elements of \texttt{dep\_var\_equ} should be the same as \texttt{loc\_arg});
the constraints are \cde differential equations and allow an immediate and
automatic replacement with the right-hand side during the calculations.

The return argument is a list of all dependent variables, like \texttt{u, psi1,
  \dots}, introduced by the command \texttt{cde\_weaklynl}, in this case:
\begin{verbatim}
dep_var_tot := {{u},{psi1,psi2,psi3},{tpsi_11,tpsi_12,tpsi_13}};
\end{verbatim}

Finally, we need a list of the two names that are used for nonlocal variables
for the two operators; in this case, since we have only one operator, we will
repeat the nonlocal variable two times:
\begin{verbatim}
nloc_arg:={{tpsi,w},{tpsi,w}};
\end{verbatim}
The Schouten bracket is readily computed:
\begin{verbatim}
sb_res:=schouten_bracket_wnl(ham1,ham2,dep_var_equ,loc_arg,nloc_arg);
\end{verbatim}
The return variable \texttt{sb\_res} contains the expression of the
three-vector $[P,P]$ after the reduction to the divergence-free normal
form. The list of variables \texttt{dep\_var\_tot} might be used to extract
coefficients of the three-vector \texttt{sb\_res} and set them to zero if
needed by the problem that is under consideration. In the mKdV case it is
obviously zero.

Let us briefly explain how \cde calculates the divergence-free normal form.

The expression~\eqref{eq:283} is a three-vector which is defined up to total
$x$-derivatives. It is processed by the software following the algorithm
developed in \cite{CLV19} in the following steps.
\begin{enumerate}
\item The command \texttt{schouten\_bracket\_wnl} makes several consistency
  checks on the input, which should be declared as above.
\item Then, the control is taken by the symbolic procedure\footnote{This is a
    \reduce concept meaning that the procedure operates on raw Rlisp data
    \hl{rather than algebraic data}, see \cite{NV14}} \texttt{sb\_wnl\_algorithm}
  which, as a first step, generates from \texttt{nloc\_arg} nonlocal variables
  which are functions of the arguments \texttt{psi1, psi2, psi3}. It is of
  utmost importance that \texttt{nloc\_arg} contains all non-identical nonlocal
  variables; using two different names for the same nonlocal variable would
  result in no simplification and wrong results.
\item The control is taken now by the symbolic procedure
  \texttt{dubrovin\_zhang\_expr}, which computes the formula~\eqref{eq:283}.
  The formula is built by blocks, one for each of the summands of~\eqref{eq:17}
  plus one for the argument $Q(\psi^2)$, with four distinct symbolic procedures
  used to that purpose, and the result is recomputed with cyclically permuted
  arguments $\psi^1$, $\psi^2$, $\psi^3$ two times, with the roles of $P$ and
  $Q$ exchanged.
\item The result is fed into a symbolic procedure,
  \texttt{nonlocal\_reduction}, that collects all terms that consist of a
  coefficient $C^{pi}$ (which can be an arbitrary function of $u^i$ and its
  derivatives) that multiplies one of the terms
  \begin{equation}
    \label{eq:4}
    \tilde{\psi}^1_\alpha \psi^2_p \partial_x^k\psi^3_i,\qquad
    \tilde{\psi}^2_\alpha \psi^3_p \partial_x^k\psi^1_i,\qquad
    \tilde{\psi}^3_\alpha \psi^1_p \partial_x^k\psi^2_i,
  \end{equation}
  with $k>0$, or similar terms with $\tilde{\chi}_\beta^a$, and replace them by
  \begin{equation}
    \label{eq:9}
    (-1)^k\partial_x^k(C^{pi}\tilde{\psi}^1_\alpha \psi^2_p)\psi^3_i
  \end{equation}
  and similar terms. In the end, the nonlocal part of the three-vector will be
  generated by
  \begin{equation}
    \label{eq:5}
    \tilde{\psi}^1_\alpha \partial_x^h\psi^2_p \psi^3_i,\qquad
    \tilde{\psi}^2_\alpha \partial_x^h\psi^3_p \psi^1_i,\qquad
    \tilde{\psi}^3_\alpha \partial_x^h\psi^1_p \psi^2_i,
  \end{equation}
  with $h\geq 0$, or similar terms with $\tilde{\chi}_\beta^a$.
\item Another symbolic procedure, \texttt{local\_reduction}, does a similar job
  on terms of the form
  \begin{equation}
    \label{eq:6}
    \partial_x^k\psi^1_j \partial_x^h\psi^2_p \partial_x^l\psi^3_i
  \end{equation}
  with respect to $\psi^3$ when $l>0$, so to obtain a local part which is of
  order $0$ with respect to $\psi^3$.
\end{enumerate}
The output of the command \texttt{schouten\_bracket\_wnl} is the three-vector
$[P,Q]$ in the normal, divergence-free form. \hl{In this example, the output is
  zero.  The elapsed time (measured by activating the Reduce switch \texttt{on
    time;}) is 30ms.  We stress that here and for the following examples we
  used CSL as the underlying Lisp system for Reduce (the other possibility
  being the PSL Lisp system, see} \cite{reduce}\hl{) on a laptop with an Intel
  i7-8565U processor.}

\subsection{The Heisenberg magnet equation}
\label{sec:heis-magn-equat}

We describe a program file to calculate the Schouten bracket of $P$ as in
\eqref{eq:101}. We will only mention the differences between the program for the
mKdV and this one. We introduce the dependent variables and the arguments of the
operator:
\begin{verbatim}
dep_var_equ:={u1,u2};
loc_arg:={psi1,psi2};
\end{verbatim}
Then we define the local part of the operator:
\begin{verbatim}
p:=(1/2)*(u1**2 + u2**2 + 1)$

mk_cdiffop(ham_l,1,{2},2);

for all psi11 let ham_l(1,1,psi11)=
  p**2*td(psi11,x) + p*(u1*u1_x + u2*u2_x)*psi11;

for all psi12 let ham_l(1,2,psi12)=
  p*(u1*u2_x - u2*u1_x)*psi12;

for all psi21 let ham_l(2,1,psi21)=
  p*(u2*u1_x - u1*u2_x)*psi21;

for all psi22 let ham_l(2,2,psi22)=
  p**2*td(psi22,x) + p*(u1*u1_x + u2*u2_x)*psi22;
\end{verbatim}
The nonlocal part contains only one `tail' vector:
\begin{verbatim}
mk_wnlop(c,w,1);
c(1,1):=1;
w(1,1):=u1_x;
w(2,1):=u2_x;
\end{verbatim}
The first and second operators in the bracket are loaded as
\begin{verbatim}
ham1:={ham_l,c,w};
ham2:=ham1;
\end{verbatim}
There is only one nonlocal variable:
\begin{verbatim}
nloc_var:={{tpsi,w,1}};
\end{verbatim}
The space of all variables and derivatives is generated by
\begin{verbatim}
dep_var_tot:=cde_weaklynl(indep_var,dep_var_equ,
  loc_arg,nloc_var,total_order);
\end{verbatim}
The list of the two names of the nonlocal variables is
\begin{verbatim}
nloc_arg:={{tpsi,w},{tpsi,w}};
\end{verbatim}
and the Schouten bracket $[P,P]=0$ is calculated by:
\begin{verbatim}
sb_res:=schouten_bracket_wnl(ham1,ham2,dep_var_equ,
  loc_arg,nloc_arg);
\end{verbatim}
The last command returns \texttt{0}; \hl{the elapsed time is 810ms, plus 40ms
  of garbage collection.}

The Schouten bracket for the second operator $Q$ \eqref{eq:21} of the Heisenberg
magnet equation \eqref{eq:13} is calculated as follows.
The operator is input as
\begin{verbatim}
p:=(1/2)*(u1**2 + u2**2 + 1);
mk_cdiffop(ham_2,1,{2},2);
for all psi11 let ham_2(1,1,psi11) = 0;
for all psi12 let ham_2(1,2,psi12) = - p**2*psi12;
for all psi21 let ham_2(2,1,psi21) = p**2*psi21;
for all psi22 let ham_2(2,2,psi22) = 0;
\end{verbatim}
The operator is local, hence we need to set its nonlocal data to $0$:
\begin{verbatim}
mk_wnlop(d,z,1);
d(1,1):=1;
z(1,1):=0;
z(2,1):=0;
\end{verbatim}
The input for the two arguments of the Schouten bracket is formed as
\begin{verbatim}
ham1:={ham_2,d,z};
ham2:=ham1;
\end{verbatim}
The jet space is generated by
\begin{verbatim}
nloc_var:={{tpsi,z,1}};
dep_var_tot:=cde_weaklynl(indep_var,dep_var_equ,
  loc_arg,nloc_var,total_order);
\end{verbatim}
and the Schouten bracket is calculated by
\begin{verbatim}
nloc_arg:={{tpsi,z},{tpsi,z}};
sb_res:=schouten_bracket_wnl(ham1,ham2,dep_var_equ,
  loc_arg,nloc_arg);
\end{verbatim}
with result $0$ \hl{in just 30ms}.

The above Hamiltonian operators for the Heisenberg magnet equation are
compatible: $[P,Q]=0$. The operators are loaded as above in a single program
file, with
\begin{verbatim}
ham1:={ham_l,c,w};
ham2:={ham_2,d,z};
\end{verbatim}
there are two nonlocal variables (even if one is trivial, it must be fed into
the program):
\begin{verbatim}
nloc_var:={{tpsi,w,1},{tchi,z,1}};
dep_var_tot:=cde_weaklynl(indep_var,dep_var_equ,
  loc_arg,nloc_var,total_order);
\end{verbatim}
The Schouten bracket $[P,Q]$ is calculated by
\begin{verbatim}
nloc_arg:={{tpsi,w},{tchi,z}};
sb_res:=schouten_bracket_wnl(ham1,ham2,dep_var_equ,
  loc_arg,nloc_arg);
\end{verbatim}
with $0$ as a result in 300ms.

\subsection{The equations of associativity}
\label{sec:equat-assoc}

Here we will compute the Schouten bracket $[P,P]$ for $P$ as in \eqref{eq:37}.
This provides an example of computation where the nonlocal part of the operator
has two distinct nonlocal summands.

After loading the variables and the arguments
\begin{verbatim}
dep_var_equ:={u1,u2,u3};
loc_arg:={psi1,psi2,psi3};
\end{verbatim}
we define the velocity matrix of the system of PDEs; in particular, the \reduce
matrix element \texttt{av(i,j)} corresponds to $\partial V^i/\partial u^j_x$ in
the system of PDEs $u^i_t = (V^i)_x$~\eqref{eq:111}.
\begin{verbatim}
% right-hand side of the system of PDEs
de:={
  u2_x + u3_x,
 - ( - 2*u1*u2*u2_x - u1*u2*u3_x - u1*u2_x*u3
     + u1_x*u2**2 + u1_x*u2*u3 - u1_x)/u1**2,
  u1_x
    }$

nc:=length(dep_var_equ)$
dv1:={u1_x,u2_x,u3_x}$

matrix av(nc,nc);
for i:=1:nc do
  for j:=1:nc do
    av(i,j):=df(part(de,i),part(dv1,j));
\end{verbatim}
Then, we define the local part of the operator by
\begin{verbatim}
mk_cdiffop(ham_l,1,{3},3);
for all i,j,psi let ham_l(i,j,psi)=
  gu1(i,j)*td(psi,x) + b(i,j)*psi;
\end{verbatim}
Here, \texttt{gu1} is a \reduce operator containing the metric of the operator
$P$ in upper indices, and \texttt{b} is a \reduce operator containing the value
of $\Gamma^{ij}_k u^k_x$ that has been separately computed (see
Subsection~\ref{sec:equat-assoc-maple}). The nonlocal part of the operator is
loaded by
\begin{verbatim}
mk_wnlop(c,w,2);
c(1,1):= - 1;
c(2,2):= - 1;
c(1,2):=0;
c(2,1):=0;
for i:=1:nc do w(i,1):=(for j:=1:nc sum av(i,j)*part(dv1,j));
w(1,2):=u1_x;
w(2,2):=u2_x;
w(3,2):=u3_x;
\end{verbatim}
The \reduce operator \texttt{c} contains the coefficients of the nonlocal
summands.  The second index of the \reduce operator \texttt{w} labels the
`tail' vector, and it corresponds to $\alpha$ in $w^i_\alpha$. In this case we
have just one operator in the bracket with three distinct nonlocal variables:
\begin{verbatim}
nloc_var:={{tpsi,w,1},{tpsi,w,2}};
\end{verbatim}
The Schouten bracket is computed by
\begin{verbatim}
nloc_arg:={{tpsi,w},{tpsi,w}};
sb_res:=schouten_bracket_wnl(ham1,ham2,dep_var_equ,
  loc_arg,nloc_arg);
\end{verbatim}
The result is \texttt{0}. \hl{Here the execution time is much higher, due to the
algebraic complexity of the example: 21330ms, with 1020ms of garbage
collection.}

\section{Mathematica implementation}
\label{sec:impl-mathematica}

We describe the implementation of the computation of the Jacobi identity for nonlocal
Poisson vertex algebras in Mathematica. To this aim, let us start by reviewing some basic facts from \cite{DSK13} (see also \cite{CLV19}).

Let $\mc V$ be an algebra of differential functions in the variables $\{u^{i}\}_{i\in I}$ (cf. \cite{BarakatSoleKac:PVAlTHEq}).
Recall from \cite{DSK13} that to a matrix pseudodifferential operator
$P=\big(P^{ij}(\partial)\big)_{i,j\in I}\in\Mat_{\ell\times\ell}\mc V((\partial^{-1}))$
we associate a map, called (nonlocal) $\lambda$-bracket,
$\{\cdot\,_\lambda\,\cdot\}_P:\,\mc V\times\mc V\to\mc V((\lambda^{-1}))$,
given by the following \emph{Master Formula}:
\begin{equation}\label{MasterFormula}
\{f_\lambda g\}_P
=
\sum_{\substack{i,j\in I \\ m,n\in\mb Z_+}}
\frac{\partial g}{\partial u_j^{(n)}}
(\lambda+\partial)^n
P^{ji}(\lambda+\partial)
(-\lambda-\partial)^m
\frac{\partial f}{\partial u_i^{(m)}}
\,\in\mc V((\lambda^{-1}))
\,.
\end{equation}
In particular, $\{{u_i}_\lambda{u_j}\}_P=P^{ji}(\lambda)$, the symbol of the $(j,i)$-entry of the matrix pseudodifferential operator $P$.
For arbitrary $P$, the $\lambda$-bracket \eqref{MasterFormula}
satisfies the following sesquilinearity conditions ($f,g\in\mc V$):
$$
\{\partial f_\lambda g\}_P=-\lambda \{f_\lambda g\}_P\,,
\qquad
\{f_\lambda \partial g\}_P=(\lambda+\partial)\{f_\lambda g\}_P\,,
$$
and left and right Leibniz rules ($f,g,h\in\mc V$):
$$
\{f_\lambda gh\}_P=\{f_\lambda g\}_Ph+\{f_\lambda h\}_Pg\,,
\qquad
\{fg_\lambda h\}_P={\{f_{\lambda+\partial} h\}_P}_\rightarrow g+{\{g_{\lambda+\partial} h\}_P}_\rightarrow f
\,.
$$
An expression of the form ${\{f_{\lambda+\partial}h\}_P}_\to g$ is interpreted as follows:
if $\{f_{\lambda}h\}_P=\sum_{n=-\infty}^Nc_n\lambda^n$,
then ${\{f_{\lambda+\partial}h\}_P}_\to g=\sum_{n=-\infty}^Nc_n(\lambda+\partial)^ng$,
where we expand $(\lambda+\partial)^n$ in non-negative powers of $\partial$.

The matrix pseudodifferential operator $P$ is skew adjoint if and only if ($f,g\in\mc V$)
\begin{equation}\label{eq:skew}
\{f_\lambda g\}_P=-\{g_{-\lambda-\partial} f\}_P\,.
\end{equation}
The RHS of the skewsymmetry condition should be interpreted as follows:
we move $-\lambda-\partial$ to the left and
we expand its powers in non-negative powers of $\partial$,
acting on the coefficients on the $\lambda$-bracket.

In general, for all $f,g,h\in\mc V$ we have that $\{f_\lambda\{g_\mu h\}\}_P\in\mc V((\lambda^{-1}))((\mu^{-1}))$.
Let
$\mc V_{\lambda,\mu}:=\mc V[[\lambda^{-1},\mu^{-1},(\lambda+\mu)^{-1}]][\lambda,\mu]$,
namely the quotient of the $\mb C[\lambda,\mu,\nu]$-module
$\mc V[[\lambda^{-1},\mu^{-1},\nu^{-1}]][\lambda,\mu,\nu]$
by the submodule
$(\nu-\lambda-\mu)\mc V[[\lambda^{-1},\mu^{-1},\nu^{-1}]]$ $[\lambda,\mu,\nu]$.
We have the natural embedding
$\iota_{\mu,\lambda}:\,\mc V_{\lambda,\mu}\hookrightarrow V((\lambda^{-1}))((\mu^{-1}))$
defined by expanding the negative powers of $\nu=\lambda+\mu$
by geometric series in the domain $|\mu|>|\lambda|$.

The $\lambda$-bracket $\{\cdot\,_\lambda\,\cdot\}_P$ defined by \eqref{MasterFormula} is called \emph{admissible} if  ($f,g,h\in\mc V$):
\begin{equation}\label{eq:admissible}
\{f_\lambda\{g_\mu h\}_P\}_P\in\mc V_{\lambda,\mu}\,,
\end{equation}
where we identify the space $\mc V_{\lambda,\mu}$
with its image in $\mc V((\lambda^{-1}))((\mu^{-1}))$ via the embedding $\iota_{\mu,\lambda}$.
It is proved in \cite{DSK13} that, if skewsymmetry \eqref{eq:skew} and admissibility \eqref{eq:admissible} conditions
hold, then
we also have $\{g_\mu\{f_\lambda h\}_P\}_P\in\mc V_{\lambda,\mu}$ and
$\{{\{f_\lambda g\}_P}_{\lambda+\mu} h\}_P\in\mc V_{\lambda,\mu}$, for every $f,g,h\in\mc V$.

A skewsymmetric and admissible $\lambda$-bracket \eqref{MasterFormula} defines a \emph{(nonlocal) Poisson vertex algebra
structure} on $\mc V$ if the following Jacobi identity is satisfied ($f,g,h\in\mc V$):
\begin{equation}\label{eq:jacobi}
\{f_\lambda\{g_\mu h\}_P\}_P-\{g_\mu\{f_\lambda h\}_P\}_P-\{{\{f_\lambda g\}_P}_{\lambda+\mu} h\}_P=0\,,
\end{equation}
where the equality is understood in the space $\mc V_{\lambda,\mu}$. In this case $P$ is called a
\emph{nonlocal Hamiltonian operator}.

Note that Jacobi identity \eqref{eq:jacobi} holds for all $f,g,h\in\mc V$
if and only if it holds for any triple of generators $u^i,u^j,u^k$, $i,j,k\in I$:
\begin{equation}\label{eq:jacobi_gen}
\{{u^i}_\lambda\{{u^j}_\mu u^k\}_P\}_P-\{{u^j}_\mu\{{u^i}_\lambda u^k\}_P\}_P
-\{{\{{u^i}_\lambda u^j\}_P}_{\lambda+\mu} u^k\}_P=0\,.
\end{equation}
Equation \eqref{eq:jacobi_gen} is equivalent to the equation $[P,P]=0$, where $[\cdot\,,\,\cdot]$ denotes the
Schouten bracket of matrix pseudodifferential operators defined in \eqref{eq:283}.

Two nonlocal Hamiltonian operators $P$ and $Q$ are called compatible if
their linear combination (or, equivalently, their sum)
is a non-local Hamiltonian operator.
According to the above discussion, this is equivalent to check that the $\lambda$-bracket
$\{\cdot\,_\lambda\,\cdot\}_{P+Q}=\{\cdot\,_\lambda\,\cdot\}_{P}+\{\cdot\,_\lambda\,\cdot\}_{Q}$ defined by equation
\eqref{MasterFormula} satisfies Jacobi identity \eqref{eq:jacobi_gen} on generators. This condition reads ($i,j,k\in I$):
\begin{equation}\label{eq:compatibility}
\begin{split}
&\{{u^i}_\lambda\{{u^j}_\mu u^k\}_P\}_Q+\{{u^i}_\lambda\{{u^j}_\mu u^k\}_Q\}_P
-\{{u^j}_\mu\{{u^i}_\lambda u^k\}_P\}_Q-\{{u^j}_\mu\{{u^i}_\lambda u^k\}_Q\}_P
\\
&-\{{\{{u^i}_\lambda u^j\}_P}_{\lambda+\mu} u^k\}_Q-\{{\{{u^i}_\lambda u^j\}_Q}_{\lambda+\mu} u^k\}_P=0\,.
\end{split}
\end{equation}
Equation \eqref{eq:compatibility} is equivalent to the equation $[P,Q]=0$.

Consider the weakly nonlocal skew-adjoint matrix pseudodifferential operators $P$ and $Q$ defined in equations \eqref{eq:18} and \eqref{eq:29}. The corresponding $\lambda$-brackets on $\mc V$, defined by equation \eqref{MasterFormula}, on a pair of
generators $u^i,u^j$, $i,j\in I$, are
\begin{align}\label{eq:lambdaP}
  & \{{u^j}_\lambda u^i\}_P =P^{ij}(\lambda)=
  B^{ij\sigma}\lambda^\sigma + c^{\alpha\beta}
  w^i_\alpha(\lambda+\partial)^{-1}w^j_\beta\,,
  \\ \label{eq:lambdaQ}
  & \{{u^j}_\lambda u^i\}_Q =Q^{ij}(\lambda)=
  C^{ij\sigma}\lambda^\sigma + d^{\alpha\beta}
  z^i_\alpha(\lambda+\partial)^{-1}z^j_\beta\,.
\end{align}
In both equations \eqref{eq:lambdaP} and \eqref{eq:lambdaQ} we should use the expansion
$$
(\lambda+\partial)^{-1}=\sum_{n\in\mb Z_+}\lambda^{-n-1}(-\partial)^n\,,
$$
where $\partial$ acts on coefficients on the right, to get elements in $\mc V((\lambda^{-1}))$.

Since the nonlocal matrix pseudodifferential operators $P$ and $Q$, defined by equations \eqref{eq:18} and \eqref{eq:29} respectively,
are rational, it follows from \cite{DSK13} that the
$\lambda$-brackets \eqref{eq:lambdaP} and \eqref{eq:lambdaQ} are admissible. In order to check that $P$ (or $Q$) is a Hamiltonian
operator we need to verify that equation \eqref{eq:jacobi_gen} holds.
To check that $[P,Q]=0$, we need to verify that equation \eqref{eq:compatibility} holds.

In Mathematica the Master Formula \eqref{MasterFormula} is implemented to compute each summand in the LHS of equations
\eqref{eq:jacobi_gen} and \eqref{eq:compatibility}. Then the algorithm described in \cite{CLV19} to write these summands in a
convenient basis of $\mc V_{\lambda,\mu}$ is implemented in order to check whether equations \eqref{eq:jacobi_gen} and \eqref{eq:compatibility} hold.
We illustrate
the implementation and the usage in the three examples considered in Section \ref{sec:impl-cdiff-cde}.

\subsection{The mKdV equation}
\label{sec:mkdv-equation-mathematica}

Consider the mKdV equation $u_t = u^3u_x + u_{xxx}$ and its nonlocal $\lambda$-bracket
(cf. equation \eqref{eq:7})
\begin{equation}\label{eq:lambda_mkdv}
\{u_\lambda u\}_P=
\lambda^3 + \frac{1}{3}\left(2\lambda+\partial\right)u^2 - \frac23u_x (\lambda+\partial)^{-1}u_x.
\end{equation}
The Hamiltonian property of $P$ is proved by checking that the Jacobi identity \eqref{eq:jacobi_gen} vanishes for the triple $u,u,u$.

We program the above calculation as follows.
\hl{The package \texttt{nlPVA.wl} (with the classical implementation of the algorithm) or \texttt{nlPVA-par.wl} (with the parallel implementation) can be either stored locally, in a path where Mathematica is able to find them, or accessed remotely from the repository \texttt{https://github.com/mcasati/nlPVA}. This makes it possible using the packages on the web-based Wolfram Cloud and Mathematica online, as well as in any desktop version of the software (from 11.0) without installation. We import the package}
\begin{sverb}
Import["https://raw.githubusercontent.com/mcasati/nlPVA/main/nlPVA.wl"]
\end{sverb}
\hl{(or the parallel implementation \texttt{nlPVA-par.wl}). The classical and parallel implementation have the same input and the same final output. The parallelization of the code is performed automatically by Mathematica using the command \texttt{Parallelize[]}.}
We initialize the package settings by introducing the name $u^i$ for the dependent variables and $x$ for the independent variable using the commands \texttt{SetGenName} and \texttt{SetVarName}, and introducing a formal
parameter $\beta$ (shadowing the formal variable $\lambda$ of the $\lambda$-bracket) using the command \texttt{SetFormalParameter}
as follows:

\begin{sverb}
SetGenName[u];
SetVarName[x];
SetFormalParameter[β];
 \end{sverb}
The formal parameter introduced will be used throughout the algorithm for internal computations and should be different from
the variable $\lambda$ that will be used for outputs.

In the case of the mKdV equation we have only one dependent variable $u:=u^1$ and the corresponding $\lambda$-bracket
\eqref{eq:lambda_mkdv} has a tail of order $1$ (using the terminology in \cite{FerAMS}). \hl{Moreover, we need to specify the highest derivative of the generators which will appear in the input with the command \texttt{SetMaxO} (its value is 5 by default, but in many examples this is unnecessarily high).} We pass this information to the program as follows:
\begin{sverb}
SetN[1];
SetTail[1];
SetMaxO[1];
\end{sverb}
Then we can load the $\lambda$-bracket \eqref{eq:lambda_mkdv}. This is done in two steps.
We start by defining its local part (the polynomial part in the variable $\lambda$):
\begin{sverb}
PLoc = {{β^3 + 2/3 gen[[1]] TD[gen[[1]]] + 2/3 gen[[1]]^2 β}};
    \end{sverb}
Note that it is important to define \texttt{PLoc} as a (in this case $1\times1$) matrix for the program to work.
In the above definition, the variable $\beta$ is used as the auxiliary variable to $\lambda$, \texttt{gen} is the vector containing the dependent variables, and the command \texttt{TD} is built-in in the
program and it corresponds to the operator of total derivative (in this case, is the total derivative with respect to $x$).
Then we enter the components of the nonlocal part of \eqref{eq:lambda_mkdv} (the term containing negative powers of $\lambda$)
by the commands
\begin{sverb}
c = {{-2/3}};
w = {{D[gen[[1]], x]}};
\end{sverb}
This is done by defining the coefficient matrix of the tail
summand $c^{\alpha\beta}$ as a matrix \texttt{c} and of $w^i_\alpha$ as a matrix \texttt{w}, cf. \eqref{eq:lambdaP}.
We combine the local and nonlocal part of the $\lambda$-bracket \eqref{eq:lambda_mkdv} using the command
\texttt{BuildBracket}:
\begin{sverb}
P = BuildBracket[PLoc, c, w, Nw];
\end{sverb}
The nonlocal part of \eqref{eq:lambda_mkdv} is understood by the program as a vector containing
the matrices $\texttt{c}$ and $\texttt{w}$, and a symbol $\texttt{Nw}$ corresponding to the term $(\lambda+\partial)^{-1}w^j_\beta$.
Note that, in the case of a purely local (respectively, purely nonlocal) $\lambda$-bracket, its nonlocal (respectively, local) part has
to be initialized to $0$.
It is possible to visualize a matrix containing the full expression of the $\lambda$-brackets among generators in an explicit form
as follows:
\begin{sverb}
GetBracket[P,λ];
\end{sverb}
The LHS of equation \eqref{eq:jacobi_gen} can be computed in this case as follows:
\begin{sverb}
CompatCheck[P, P, {λ,μ,ν}];
\end{sverb}
The variables $\lambda$ and $\mu$ appearing as input correspond to the variables appearing in the Jacobi identity
\eqref{eq:jacobi_gen}. The variable $\nu$ is hidden from that identity since in the definition of $\mc V_{\lambda,\mu}$
in Section \ref{sec:impl-mathematica} we set $\nu=\lambda+\mu$.

The command \texttt{CompatCheck} implements the computation of the LHS of Jacobi identity \eqref{eq:jacobi_gen} as follows.
First, it computes all the three terms of that identity by using the Master Formula \eqref{MasterFormula}. This implementation
is the same used in the Mathematica package \texttt{MasterPVA}, see \cite{casati16:_master_walg}.
Then, it applies a normalization procedure consisting in writing each summand in  a convenient basis of $\mc V_{\lambda,\mu}$
implementing the algorithm described in \cite{CLV19}. After expressing each summand in the same basis, the program carries an easy computation of the LHS of equations \eqref{eq:jacobi_gen}. \hl{The running time of the procedure is, in general, heavily dependant on the maximal order of the generators' derivatives. It is worthy noticing, moreover, that the first time the command \texttt{CompatCheck} is used in the parallel implementation of the algorithm it takes significantly longer because Mathematica needs to initialize the additional kernels. It is possible to avoid this using the function \texttt{LaunchKernels[]} beforehand. The time required is computed by the Mathematica function \texttt{RepeatedTiming[]}. In a Mathematica 11.3 worksheet running on a laptop with an Intel i7-6500U processor and two available Mathematica kernels, the classical implementation takes $0.020$s, while the parallel one $0.028$s}

\subsection{The Heisenberg magnet equation}
\label{sec:heis-magn-equat-mathematica}
Let $P(\partial)$ be the matrix pseudodifferential operator defined in \eqref{eq:101}. The corresponding $\lambda$-brackets
among generators are given by the following matrix equation
\begin{equation}
  \label{eq:10-lambda}
  \begin{split}
&\begin{pmatrix}
\{{u^1}_\lambda u^1\}_P&\{{u^2}_\lambda u^1\}_P
\\
\{{u^1}_\lambda u^2\}_P&\{{u^2}_\lambda u^2\}_P
\end{pmatrix}
=P(\lambda)
\\
& =
  \begin{pmatrix}
    f (\lambda + \partial) f +    u^1_x (\lambda+\partial)^{-1} u^1_x& f(u^1 u^2_x - u^2 u^1_x)+u^1_x (\lambda+\partial)^{-1} u^2_x
    \\
    f(    u^2 u^1_x - u^1 u^2_x)+u^2_x (\lambda+\partial)^{-1} u^1_x & f(\lambda+\partial)f+u^2_x (\lambda+\partial)^{-1} u^2_x
  \end{pmatrix}
\,,
\end{split}
\end{equation}
where $f=(1/2)((u^1)^2 + (u^2)^2 + 1)$.

Let us explain how to check that $P$ is a nonlocal Hamiltonian operator.
After loading the package and initializing the package settings as in Section \ref{sec:mkdv-equation-mathematica} we introduce
the number of dependent variables
\begin{sverb}
SetN[2];
\end{sverb}
This command builds up a vector \texttt{gen} with two components corresponding to the dependent variables $u^1$ and $u^2$. \hl{The maximal order of derivatives of the generators in the two operators is 1 as in the previous example, so we do not need to reassign the value of \texttt{MaxO} (we can check that \texttt{GetMaxO[]} outputs 1).}
Then we define the local part of the operator:
\begin{sverb}
f = 1/2(gen[[1]]^2+gen[[2]]^2+1);
PLoc = {{f^2 β+f TD[P],
  f (gen[[1]] TD[gen[[2]]] - gen[[2]] TD[gen[[1]]])},
 {f (gen[[2]] TD[gen[[1]]] - gen[[1]] TD[gen[[2]]]),
  f^2 β+f TD[f]}} ;
\end{sverb}
The components of the nonlocal part are defined as follows:
\begin{sverb}
c = {{1}};
w = {{TD[gen[[1]]], TD[gen[[2]]]}};
\end{sverb}
Then we combine the local and nonlocal parts of the $\lambda$-brackets \eqref{eq:10-lambda}:
\begin{sverb}
P = BuildBracket[PLoc, c, w, Nw];
\end{sverb}
To check that $P$ satisfies Jacobi identity we use
\begin{sverb}
CompatCheck[P, P, {λ, μ, ν}];
\end{sverb}
to compute the LHS of equation \eqref{eq:jacobi_gen}. Note that, by default,
this commands prints as output the results of the partial steps of this
computation corresponding to the different choices of the indices $i,j,k\in I$. \hl{This computation returns a set of four zeroes, corresponding to the four sets of indices (1,1,1), (1,1,2), (1,2,2), (2,2,2) and it is performed in $0.23$s using the classical implementation and $0.15$s using the parallel implementation.}

Let $Q()$ denote the second Hamiltonian operator of the Heisenberg magnet equation defined in equation \eqref{eq:21}.
The corresponding $\lambda$-brackets among generators are
\begin{equation}
  \label{eq:21-lambda}
  \begin{split}
&\begin{pmatrix}
\{{u^1}_\lambda u^1\}_Q&\{{u^2}_\lambda u^1\}_Q
\\
\{{u^1}_\lambda u^2\}_Q&\{{u^2}_\lambda u^2\}_Q
\end{pmatrix}
=Q(\lambda)=\begin{pmatrix}0&-f^2\\f^2&0
  \end{pmatrix}
\,.
\end{split}
\end{equation}
The compatibility condition between the non-local $\lambda$-bracket \eqref{eq:10-lambda} and the local $\lambda$-bracket
\eqref{eq:21-lambda} is checked as follows.
First, we define the $\lambda$-brackets \eqref{eq:21-lambda} by
\begin{sverb}
d = {{0, 0}, {0, 0}};
y = {{0, 0, 0}, {0, 0, 0}};
Q = BuildBracket[Qloc, d, y, Nz];
\end{sverb}
where \texttt{Qloc} is the matrix \eqref{eq:21-lambda} separately introduced in Mathematica. As previously mentioned, in the case that the
$\lambda$-brackets are local we still need to define their nonlocal tail to be zero for the program to work.
Then, using the command
\begin{sverb}
CompatCheck[P, Q, {λ, μ, ν}];
\end{sverb}
one can check that the compatibility condition \eqref{eq:compatibility} is  satisfied. \hl{The running time of this function is
$0.11$s with the classical implementation and $0.081$s with the parallel one.}
\subsection{The equations of associativity}
\label{sec:equat-assoc-mathematica}

Consider the $\lambda$-brackets on generators associated to the
weakly nonlocal matrix pseudodifferential operator $P(\partial)$ defined in \eqref{eq:37} ($i,j=1,2,3$)
\begin{equation}
  \label{eq:37-mathematica}
\{{u^j}_\lambda u^i\}_P = g^{ij}\lambda + \Gamma^{ij}_{k}u^k_x
  +\alpha_1 \pd{V^i}{u^q}u^q_x(\lambda+\partial)^{-1}\pd{V^j}{u^p}u^p_x
  +\gamma_1 u^i_x(\lambda+\partial)^{-1}u^j_x,
\end{equation}
where the metric $g^{ij}$ is defined in \eqref{eq:38}, the Christoffel symbols $\Gamma^{ij}_k$ are defined in
\eqref{eq:82}, \eqref{eq:83} and \eqref{eq:84},
and the value of the constants is $\alpha_1=-1$, $\gamma_1=-1$.

Let us show the required steps to check that $P$ is a Hamiltonian operator. After loading the package and initializing its settings as in Section \ref{sec:mkdv-equation-mathematica} we introduce the number of dependent variables
and the length of the tail (which in this case is $2$)
\begin{sverb}
SetN[3];
SetTail[2];
\end{sverb}
We define the local part of the $\lambda$-brackets \eqref{eq:37-mathematica} by
\begin{sverb}
PLoc = g β  + Γ1 TD[gen[[1]]] + Γ2 TD[gen[[2]]]
       + Γ3 TD[gen[[3]]];
\end{sverb}
Here, \texttt{g} is the matrix in \eqref{eq:38}, \texttt{Γ1} the matrix in \eqref{eq:82},
\texttt{Γ2} the matrix in \eqref{eq:83} and \texttt{Γ3} the matrix in \eqref{eq:84},
which were previously separately introduced in Mathematica.
The nonlocal part
of the operator is loaded by
\begin{sverb}
c = {{-1, 0}, {0, -1}};
w = {Table[Sum[D[V[[i]], gen[[s]]] TD[gen[[s]]],
    {s, $d}], {i, $d}],
    Table[TD[gen[[i]]], {i, $d}]};
  \end{sverb}
  The matrix \texttt{c} contains the coefficients of the nonlocal summands.
  The second index of matrix \texttt{w} labels the tail vector. The vector
  \texttt{V} contains the components $V^i$, needed to compute the velocity
  matrix $\partial V^i/\partial u^j$ of the system of PDEs \eqref{eq:111}.  We
  combine the local and nonlocal part of the $\lambda$-brackets
  \eqref{eq:37-mathematica} as follows
\begin{sverb}
P = BuildBracket[PLoc, c, w, Nw];
\end{sverb}
We check that Jacobi identity \eqref{eq:jacobi_gen} holds using the command
\begin{sverb}
CompatCheck[P, P, {ρ, σ, τ}];
\end{sverb}
(the choice of the output formal variables $\rho$, $\sigma$ and $\tau$ does not
affect the results provided that they are all distinct). \hl{The function outputs the expected list of ten zeroes (corresponding to independent entries of the Jacobi identity for a 3-components system) in $2.8$s when using the classical implementation and $1.8$s with the parallel one.}

\subsection*{Acknowledgments}  We would like to thank
A. Carbotti, E.V. Ferapontov, S. Perletti for scientific discussions.

This research has been funded by the Dept. of Mathematics and Physics ``E. De
Giorgi'' of the Universit\`a del Salento, Istituto Naz. di Fisica Nucleare
IS-CSN4 \emph{Mathematical Methods of Nonlinear Physics}, GNFM of Istituto
Nazionale di Alta Matematica.  P.~L. is supported by MIUR - FFABR funds 2017
and by funds of H2020-MSCA-RISE-2017 Project No. 778010 IPaDEGAN.


\end{document}